\documentclass[
showpacs,
aps,prd,
nofootinbib,
superscriptaddress,
tightenlines,
amsmath,
amssymb
]{revtex4}
\usepackage{bm}
\usepackage{color,graphicx}
\usepackage{amsmath}
\usepackage{amssymb}

\begin{document}
\title{Exploring Nucleon Structure with the Self-Organizing Maps Algorithm} 

\author{Evan M. Askanazi} 
\email{ema9u@virginia.edu}
\affiliation{Department of Physics, University of Virginia, Charlottesville, VA 22901, USA.}

\author{Katherine A. Holcomb} 
\email{kholcomb@virginia.edu}
\affiliation{University of Virginia Alliance for Computational Science and Engineering, University of Virginia, Charlottesville, VA 22901, USA.}

\author{Simonetta Liuti} 
\email{sl4y@virginia.edu}
\affiliation{Department of Physics, University of Virginia, Charlottesville, VA 22901, USA.}
\affiliation{Laboratori Nazionali di Frascati, INFN, Frascati, Italy}

\pacs{13.60.Hb, 13.40.Gp, 24.85.+p}

\begin{abstract}
We discuss the application of an alternative type of neural network, the Self-Organizing Map to extract parton distribution functions from various hard scattering processes. 
\end{abstract}
\maketitle

\section{Introduction}
%%%% SOMs
Artificial Neural Networks (ANNs) are an algorithm model inspired by the human brain's capacity to perform the complex operations of learning, memorizing and generalizing.
The goal of ANNs is, however, to solve objective problems which are by far less complex relatively to the human brain
capabilities.  Their basic units are sets of nodes that are defined as neurons because they can take sets of input parameters and either retain or communicate a signal in a similar way to how signals propagate from one neuron to the other as the neurons  get activated/fire. This procedure is defined via learning algorithms. Its main success is in that it allows one to estimate non-linear functions of input data. 

\begin{figure}
\includegraphics[width=8cm]{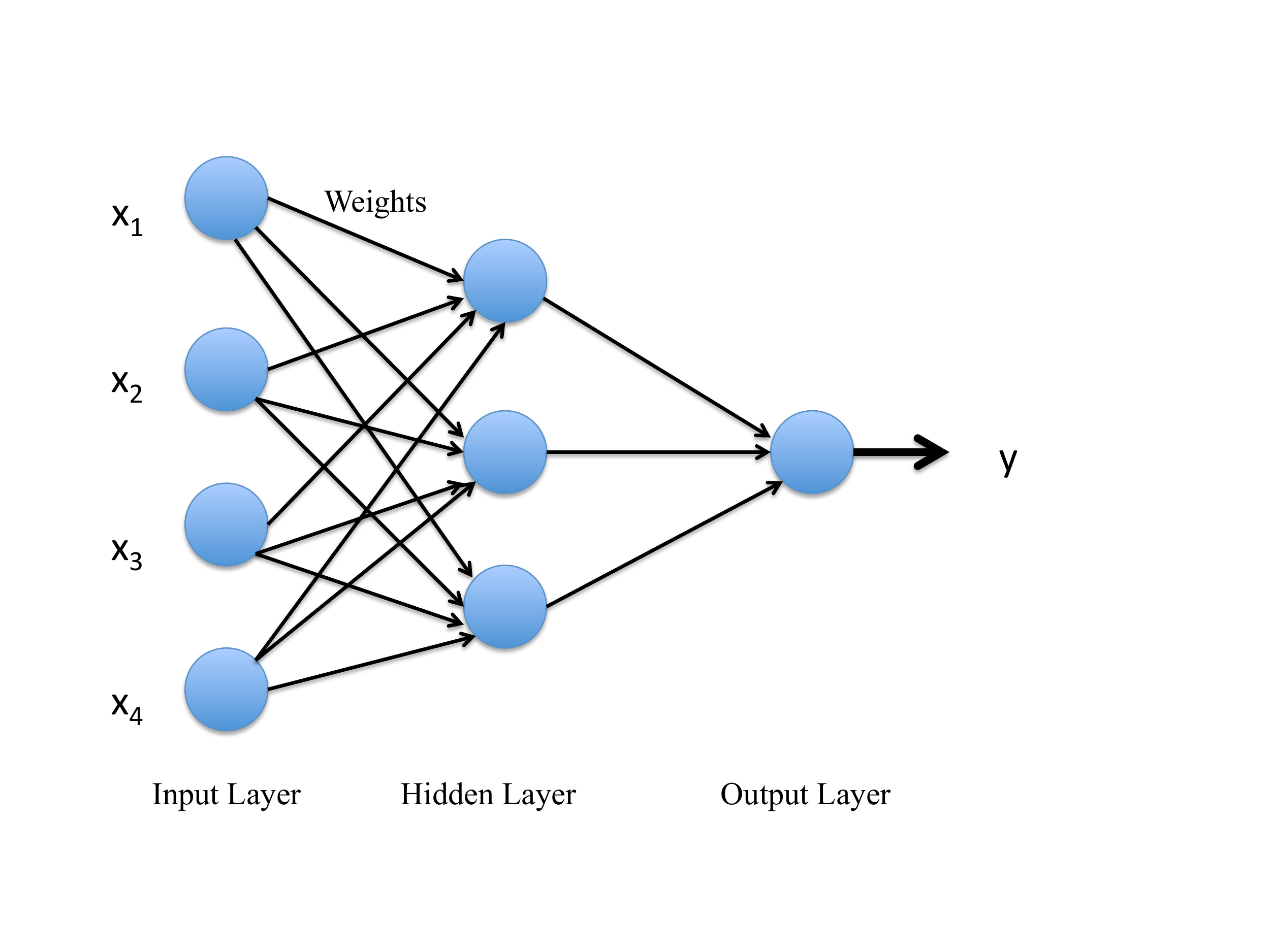}
\includegraphics[width=8cm]{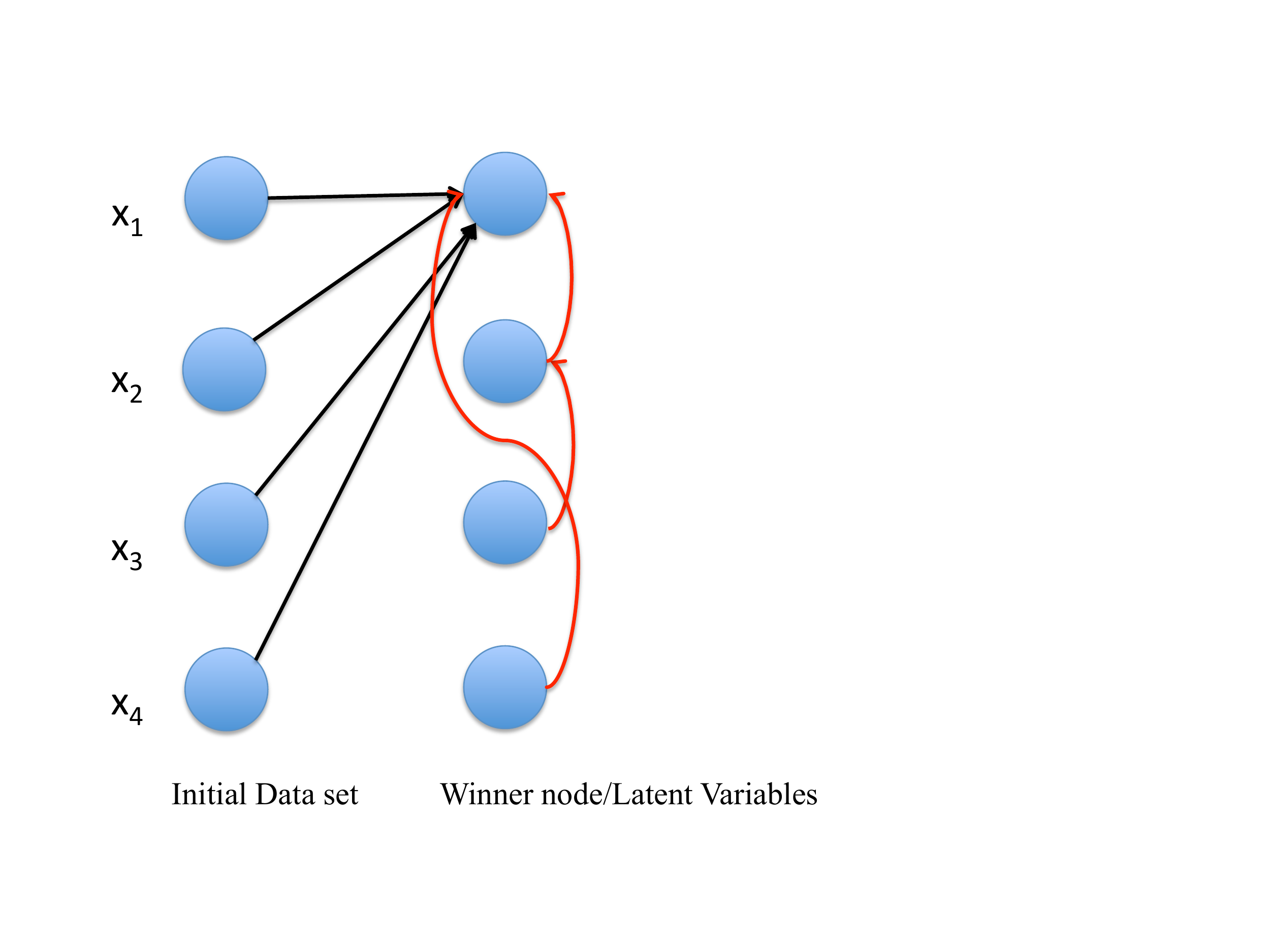}
\caption{Artificial neural networks based on supervised learning (left), and unsupervised learning (right).}
\label{ANN:fig1}
\end{figure}

ANNs consist of a set of initial data forming an input layer, a process by which the input data are evolved and trained (hidden layers), and a resulting set of output data, or the output layer 
(Figure \ref{ANN:fig1}).
%%%%
%%%%
Researchers currently utilize ANNs in data visualization, function modeling and approximating values of functions, data processing, robotics and control engineering. In the past twenty years ANNs have also established their role as a remarkable computational tool in high energy, nuclear and computational physics analyses. Important applications have been developed, for instance, as classification methods for off-line jet identification and tracking, on-line process control or event trigger and mass reconstruction, and optimization techniques in {\it e.g.}  track finding and classification \cite{Pet}. 

Neural networks are used for modeling in these fields of physics because they can utilize the principle of learning
with regards to data sets and models. The learning can be {\it supervised} or {\it unsupervised}. 

In supervised learning, the input data, $x_i$ in Fig.\ref{ANN:fig1} (left), determine the final output: a set of training examples is first formed by matching the input data (vectors) to the desired output; the learning algorithm, by proceeding through the training examples, produces a function that generalizes the mapping of the input to output vectors; this function allows one to predict correctly an output for subsequently given input data. 

\begin{figure}
\includegraphics[width=8cm]{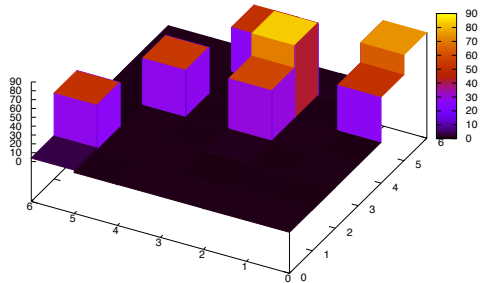}
\caption{Example of SOM. Each cell in the SOM represents $\chi^2$ values obtained in a fitting procedure. One can clearly see how the data cluster in low $\chi^2$ values (dark areas) and high $\chi^2$ values (light areas).}
\label{ANN:fig3}
\end{figure}

In unsupervised learning, Fig.\ref{ANN:fig1} (right), the training proceeds with no matching between the initial vectors and the output: a sample of initial data forms the training set;  these are compared to a distinct set of input data, and clustered according to similarities. The connection between the clustering of the initial data and the output is regulated by latent variables, thus extending and generalizing, in this respect, the definition of the learning function. 
The unsupervised learning-type algorithms allow one to uncover correlations among the input data features. The set of patterns and clustering properties seen among the data is called {\it self-organization}  \cite{Kohonen}. 
A most important aspect of self-organizing algorithms is in their ability to project high dimensional {\it input} data onto lower dimensional representations while preserving the topological features present in the {\it training} data. 
Because results using unsupervised learning are most often represented as 2D geometrical configurations 
where a neighborhood of nodes get simultaneously updated while reproducing the clustering of the data's features, 
the new algorithm is defined as a ``map", or a Self-Organizing Map (SOM).  In Figure \ref{ANN:fig3} we show an example of a SOM where each cell corresponds to $\chi^2$ values from a fitting procedure. From the figure one can clearly distinguish clusters of low vs. high $\chi^2$ data. By examining the content of each cell, {\it i.e.} looking at the curves that produced the given $\chi^2$ value for the cell, one can reconstruct the features of the curves that lead to the $\chi^2$ distribution in a more efficient way than by using any standard/back propagation based method.  

When new data are processed through SOMs they will be compared and matched through the same similarity criterion, using the data features which were determined in the training.  This aspect makes SOMs a valuable method for searching for patterns and non-linear correlations in multivariable dependent datasets. 

%there is a set of data and a neural network that creates a final output without another set of data used to supervise the mapping. 
%
%%%%
%%%%

Supervised learning based ANNs have been used to extract Parton Distribution Functions (PDFs) from high energy scattering processes \cite{Ball.2010,DelDebbio.2007,Forte.2002}.
The set of PDFs parametrizations called Neural Networks PDFs (NNPDFs), were obtained from global fits to a comprehensive set of deep inelastic scattering and collider data.
The NNPDF analysis differs from standard global analyses in that they avoid the bias that is associated with giving an analytical parametric functional form for the PDFs. 
This can be alternatively replaced by a functional form produced by a neural network with a redundant set of 37 free parameters which are given by the ANN's weights. 
The PDFs obtained this way are 
fitted to the experimental data, and their $\chi^2$ is minimized, by using a genetic algorithm approach \cite{DelDebbio.2007}. 
%Attention must be paid at this stage to the statistical aspects of the approach, and in particular of the synthetic data used in the training.
Several estimators where studied in \cite{Ball.2010,DelDebbio.2007,Forte.2002} 
to assess the quality of the ANN training. These include a ``convergence condition", or ``stopping criterion", which marks the duration of the training phase by the onset of a stage where the neural network begins to ``overlearn", or to reproduce the statistical fluctuations of the data rather than their underlying physical law.
 
%In a nutshell, what distinguishes NNPDFs from other methods is, besides an improved statistical analysis, their automated minimization procedure.
The advantage of having eliminated the user's bias or the initial model dependence in extracting PDFs from experimental data is in that
no theoretical assumption is used for determining the shape of the parametrization, rather the Bjorken $x$ dependence of the PDFs is inferred directly from the data
in terms of smoothed out curves. 
The inherent unbiasedness of this approach is a consequence of supervised learning meaning that the approach will strictly work quantitatively only in kinematical regions 
where experimental data exist. NNPDFs do not work efficiently, or they have a low performance, for extrapolating or predicting the behavior of 
the parton distribution functions. 

The ability to extrapolate and predict is however, often a desirable feature in the performance of a fit to high energy physics data.
In particular, the new generation of experiments 
that include both unpolarized and polarized semi-inclusive and exclusive high energy  scattering processes off hadronic targets (see Refs.\cite{Diehl_rev,BelRad} for a review) 
is appreciably more challenging to analyze than inclusive scattering. These experiments involve a larger number of observables and kinematical variables, 
and their kinematical coverage is by far less complete than in the inclusive sector. 
A parametrization that is free from the bias implicit in choosing an initial form, and at the same time it enables us to extrapolate where data are not available, becomes a much sought after tool in this sector. 

SOMs can provide a complementary algorithm to NNPDFs.

In our SOM-based analysis we are inspired by pioneering work also using SOMs for the analysis of high energy physics data introduced in Ref.\cite{Lonn}.  
A new parametrization PDFs parametrization, SOMPDF, was presented in Ref.\cite{Carnahan}. The initial results
were aimed at proving the viability of the method as a fitting procedure, by successfully generating $\chi^2$ values that decreased
with the increasing number of iterations. These studies did not focus, however, on the specific clustering properties of the map.

Significant restructuring of the original code was subsequently introduced in Ref.\cite{Ask} allowing us to develop a quantitative analysis of PDFs 
that goes beyond the original goal of ``developing and observing the unconventional usage of the SOM as a part of an optimization algorithm" \cite{Carnahan}.
Two important changes were introduced with respect to the original code:
\footnote{Restructuring the code included re-defining its architecture using Fortran 90. A usable version of the code will be soon available on an upcoming website.}
 {\it i)} a new initialization procedure was defined where the initial set of PDFs is taken as bundles of parametric forms whose parameters get randomly varied. This replaced the operation of varying the values of PDFs at each value of $x$ and $Q^2$ of the data, and it allowed us to obtain smooth or continuous solutions in the initial stages of the fit; 
{\it ii)} a fully quantitative error analysis was applied to our extracted PDFs.

The new code is now sufficiently flexible so that it can be applied to different multivariable dependent observables, including, in particular,  the structure functions for deeply virtual exclusive scattering (the Generalized Parton Distributions, GPDs), semi-inclusive scattering (the Transverse Momentum Distributions, TMDs), and other related processes.
Since on on side these processes are characterized by a larger number of kinematical variables and observables as compared to inclusive scattering, 
and, on the other, the extraction of hadronic structure from data depends on several model dependent theoretical assumptions, it is important to develop an ANN-based algorithm that can 
handle feature/pattern recognition problems. 

The PDFs analysis performed in \cite{Ask} did not directly make use of the pattern recognition features of SOMs. Here we present results on the performance of the new SOMPDF code  specifically for this aspect, by focusing on the extraction of PDFs at large Bjorken $x$.  This constitutes an intermediate step before addressing the extraction of more complicated objects -- the GPDs and TMDs -- from data. The large $x$ behavior of the proton's structure functions from inclusive scattering is difficult to track down since many effects, ranging from Target Mass Corrections (TMCs), Large $x$ resummation (LxR), and Higher Twists (HTs), affect the extraction of the PDFs from data by modifying the $Q^2$ dependence of the structure functions from pure Perturbative QCD (PQCD) evolution.    

This paper is organized as follows: in Section \ref{sec2} we give an overview the SOMPDF method; in Section \ref{sec3} we review the results of our PDFs analysis, and we describe in detail the clustering performance for a specific pattern recognition problem at large $x$; 
in Section \ref{sec4} we introduce and discuss the extension to multi-variable distributions, and in Section \ref{sec5} we draw our conclusions. 

%%%%%%%%%%
%%%%%%%%%%
%%%%%%%%%% SECTION II
%%%%%%%%%%
\section{Overview of the SOMPDF Approach}
\label{sec2}
The basic SOM algorithm is best described as a {\em non linear} extension of Principal Component Analysis (PCA) \cite{PCA}. 
In PCA one applies an orthogonal transformation to convert a set of data that are possibly correlated into sets of values that are linearly uncorrelated. The new sets of values constitute the principal components. 
The first principal component exhibits the largest variance, {\it i.e.} it is a straight line that minimizes the sum of the squared distances from the data points (least squares approximation) of the data set by a line. The second principal components is obtained by subtracting from the original data vectors their projections onto the first principal component and by finding a new straight line approximation. The procedure is applied to the following components recursively.
PCA is useful for interpreting the behavior of high dimensional data because, by allowing one to represent the dominant data sets in a linear form, 
and by simultaneously discarding the sub-dominant components, PCA can reduce the number of dimensions of the problem. 
However, PCA cannot account for nonlinear relationships
among data. Furthermore, it has  poor visualization properties in cases where more than two dimensions are important.

The essential feature that sets the SOM algorithm apart from PCA and similar data reduction methods is that the lines resulting from PCA can be effectively replaced by lower dimensional manifolds in the SOM method. Because of their flexibility, these can catch features of the data that the PCA would not.
In addition, SOMs have enhanced visualization features to represent higher dimensional data, while visualization for more than two or three components is unrealistic for PCA \cite{Haykin}. 

Another attractive feature of  SOMs is that they are particularly relevant algorithms in systems theory, as they model the emergence of a collective ordering in a composite system through the competition of its constituents. 
We can foresee a number of future applications to complex nuclear and high energy data using this aspect of the SOM method.  The development of our method is in fact motivated by a similar problematic to the one addressed in Ref.\cite{Ireland} where the type an number of  measurements which impact the extraction of physical observables  in a specific example was estimated by defining and detecting variations in the Shannon entropy, or 
the information content from the measurements. The specific example considered in \cite{Ireland} was polarized photoproduction of pseudoscalar mesons.
% 
%SOMs are an unsupervised network that uses a {\it two-dimensional} grid for mapping and fitting a {\it multi-dimensional} set of input data.
%%
%\footnote{The algorithm is defined as a ÒmapÓ because results using unsupervised learning are most often represented as 2D geometrical configurations.} 
%%

In our approach we combine a SOM algorithm to project high dimensional input data onto lower dimensions representations while preserving the topological features present in the training data and a Genetic Algorithm (GA) to fit input models to final data sets generated from high energy scattering experiments. 

%The method by which the SOM works allows us to use experiments conducted for a given physics process
The data used  to train the SOM are sets of PDFs obtained by varying randomly various pre-defined parametric forms in such a way that a sufficient degree of random variation is ensured while keeping the the program efficient by spanning a finite set of varied PDFs. 
From the PDFs one constructs the measurable quantities, or structure functions of deep inelastic processes in QCD. In this paper we focus on (see Ref. \cite{EFP} for detailed definitions),
\begin{eqnarray}
F_2(x,Q^2) & = & x \sum_{q,\bar{q}} e_q^2 \int_x^1 \frac{dy}{y} C_q(y,\alpha_S) \, q\left(\frac{x}{y},Q^2\right)   
+  x  \int_x^1 \frac{dy}{y} C_g(y,\alpha_S) \, g\left(\frac{x}{y},Q^2\right),
\end{eqnarray}
where $q =u,d,s,c,b$; $e_q$ is the quark charge; $g$ is the gluons distribution; the intrinsic heavy quarks component start at $Q^2 \geq m_{c(b)}^2$; $C_{q,g}$are the coefficient functions in the $\overline{\rm MS}$ scheme, and $\alpha_S(Q^2)$ is the running strong coupling.  

The observables of interest in this paper are the DIS proton and neutron electromagnetic structure functions, $F_2^{p,n}$. 
%The neutron structure function is extracted from deuteron experimental data using  $F_2^d=F_2^p+F_2^n$. 
%The data sets are described in Section \ref{sec3}.
PDFs have been extracted at NLO and NNLO from deep
inelastic lepton-nucleon scattering data, and  to NLO from collider measurements, in an ever growing range of $x$ and $Q^2$ -- from the large $x$ multi-GeV region
fixed target measurements at Jefferson Lab \cite{Acc1,Acc2,CouLiu,BiaFanLiu}, to the range of LHC precision measurements of $W^\pm$, $t\bar{t}$.
Several groups have determined the parameterizations for the unpolarized PDFs. A summary of all current PDFs parametrizations and their uncertainties is given in \cite{bench}. 

Throughout this paper 
we will compare our results with  global analyses performed by
CT \cite{CT10},  ABM \cite{ABM},  and CJ \cite{Acc1,Acc2}.  CT, ABM, CJ use a parametric form for the PDFs with several free parameters per parton distribution type, 
at an input scale, $Q_o^2$,  which varies for the different fitting forms.
\vspace{0.3cm}
The SOMPDF approach consists of the following basic components:

\vspace{0.3cm}
$\bullet$ Initialization

\vspace{0.3cm}
$\bullet$ Training

\vspace{0.3cm}
$\bullet$ Mapping

\vspace{0.3cm}
$ \bullet$ Genetic Algorithm

%%%%
\begin{figure}
\includegraphics[width=8cm]{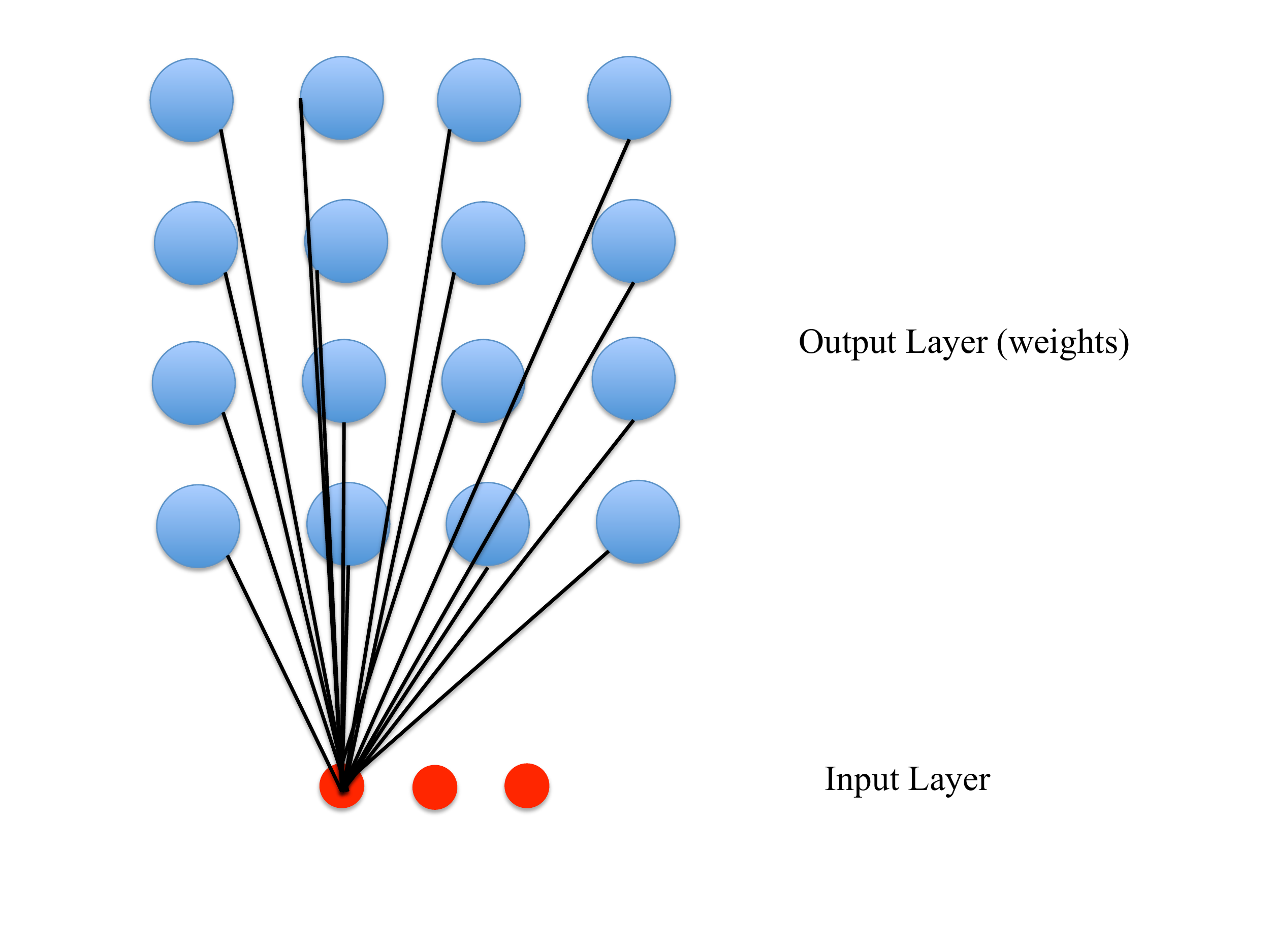}
\caption{SOM initialization. In this simple case, we consider a $N \times N$ map, $N=4$, with weight vectors of dimension $n=3$. The first element is placed on the map as shown in the figure. The other elements undergo the same procedure.}
\label{SOM:fig2}
\end{figure}
%%%%
%% Nodes
%The SOM is formed by a two dimensional $n\times n$ grid of neurons, or nodes.\footnote{In our case we choose a square map, other topologies are possible \cite{Kohonen}.}
% Input Vector
%
\subsection{Initialization}
\label{sec2A}
In the initialization phase we take a set of $n$-dimensional input data vectors as the set of data to be processed and we refer to them  as {\it code vectors}.  We form an $N \times N$ map where each cell, or node, $i=1...N^2$ represents a neuron. %and uses a competitive learning process to train the data. 
We then consider a second set of vectors -- the weight vectors $V_i$ -- with same dimension $n$ as the input vectors,
\[ V_i = [v_i^{(1)}, ..., v_i^{(n)} ].  \] 
The set $V_i$ represents a stimulus to the neurons.    
Each node $i$ is made correspond to the weights of dimension $n$: 
$V_i$ are given spatial coordinates, {\it i.e.} one defines the geometry/topology of a 2D map
that gets populated randomly with $V_i$.  In our case each one of these vectors consists of a randomized value of the type of data that is to be represented.  We 
refer to these initial vectors as {\it map vectors}.
Figure \ref{SOM:fig2} illustrates the initialization step schematically.

In our case, the vectors are sets of candidate PDFs, $V_i\{{\bf x}_m\}  \rightarrow f_i(\{{\bf x}_m\},Q_o^2)$ ($f=q, \bar{q}, g$, $i$, is the map index) which are randomly generated to form an initial bundle/envelope;
$\{{\bf x}_m\}$ is a vector of $m$ Bjorken $x$ values; each PDF is evaluated at the initial scale, $Q_o^2$.  

Both sets of code vectors and map vectors undergo NLO PQCD evolution to  the $Q^2$ of the ``external" experimental data as the competitive process that we describe below is started.  
Note, however, that at this stage the experimental data are not used: the network self-organizes itself, in an unsupervised way, through latent variables, displaying natural patterns in the data.
External data are used only later, in the GA stage of the fitting procedure. 
%In view of this final step, the PDFs are evolved to the $Q^2$ of the various experimental data sets. 

The PDFs envelope is constructed in such a way that  the PDFs behavior in $x$ varies sufficiently randomly to meet the criterion of unbiasedness, and at the same time it somewhat loosely follows the experimental data. The latter criteria were met by performing a set of ``experiments" in a detailed study in Ref.\cite{Ask}. For instance, we found that using the NLO parametric form \cite{JR1,JR2} at $Q_o^2 = 0.34$ GeV$^2$,
%
%\begin{subequations}
\begin{eqnarray}
\label{qval}
x f_j & = & A_j \, x^{B^j_1} (1-x)^{B^j_2} (1 + C^j_1 \sqrt{x} + C_2^j \, x)
\end{eqnarray} 
%\end{subequations}
where $j=u_v, d_v, \bar{u}, \bar{d}, s, c, g$,  several parameters had very little impact on the range of the overall result when they were varied within 
reasonable ranges, while others  easily drove the computed PDF to overflow without a compensating improvement in the envelope of the resulting PDFs.  
In this example the parameters that were varied so as to accomplish the best 
bracketing of our entire set of experimental data were instead $A_i$, and
all the exponential parameters, namely  $B^j_1$, $B^j_2$.
In order to compute a vector for the SOM, we multiplied each parameter, $P\equiv(A_j, B^j_1, B^j_2)$, by a random variable from a normal distribution with a mean of $1$ 
and a specified standard deviation, 
\begin{equation}
\label{random1}
P^{new} = P \pm \Delta P, 
\end{equation}
where $\Delta P$ corresponds to the distribution with $\sigma = 0.1$.
This procedure was repeated using other parametric forms {\it e.g.} \cite{ABM,CT10} and the final envelope was generated by mixing results from each form according to,  
\begin{equation}
\label{random2}
f_j^{env} = C_1 f^1_j + C_2 f^2_j + C_3 f^3_j,
\end{equation}
where each coefficient, $C_k$, $k=1, 2, 3$, is a uniform random number; $f^k_j$ are PDF forms which are randomized using Eq.(\ref{random1}).  
To mix the three parametric forms they have to be brought to a common initial $Q^2$. We accomplished this by letting the parameters depend on the variable,
\begin{equation}
s = \log \frac{\log Q^2/\Lambda^2}{\log Q_o^2/\Lambda^2} 
\end{equation}
as suggested in early PDF analyses (see {\it e.g.} \cite{GRV} and references therein).  
We then imposed the constraints  from the baryon number and momentum sum rules on each PDF,
\begin{eqnarray}
\label{baryon}
&& \int_0^1 dx \, u_v  =  2, \;\;\;\;\;\; \int_0^1 dx \, d_v = 1 \\
\label{momentum}
&& \int_0^1 dx x \, \left[(u_v + 2 \bar{u}) + \left( d_v + 2 \bar{d} \right) + 2 \bar{s} + 2 c + g \right ] = 1
\end{eqnarray}
Notice that the capacity of the PDFs generated above to properly fit existing data, are important for constructing the envelope since at this stage we just require that the bundle of input functions can encompass the data.
This step of our analysis requires a very careful study of the PDFs behavior since it is not always obvious which parameters should be chosen in the baseline parametrization formulae, Eq.(\ref{random2}), to bracket newer data.

\begin{figure}
%\includegraphics[width=7.cm]{F2envelopeint.pdf}
%\hspace{-4.5cm}
\includegraphics[width=7.cm]{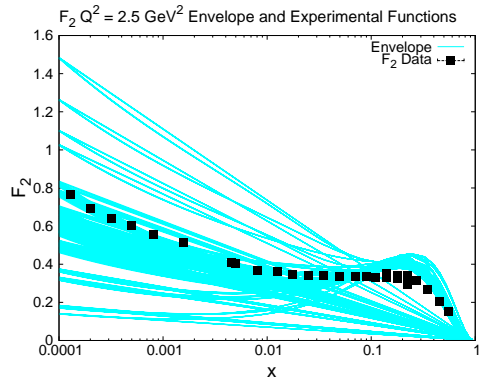}
\caption{(Color online) Example of an envelope constructed for the structure $F_2$ in the SOMPDF initialization stage at $Q^2=2.5$ GeV$^2$. The experimental data are from the set listed in Section \ref{sec3}. 
%On the LHS we show the spread of curves for the initial GA step; on the RHS we show the same spread but at the final GA step (in this case for a total number of iterations of the GA, $N_{MAX}=250$).
}
\label{fig:envelope}
\end{figure}

An example of an envelope is given in Figure \ref{fig:envelope}. 
Each generated PDF now contains a mixture of up to three different $x$ and $Q^2$ dependent parametric forms whose parameters were varied in order to assure both randomness and the observance of physical constraints.  A set of these makes up a vector for a particular cell; the number generated for each cell is a parameter of the code.  Once the envelope is formed, we proceed with the training and the final part of our analysis, or the implementation of the GA. In fact, the resulting envelope of PDFs that effectively wrapped around the experimental data from above and below was constructed so that it minimizes the  bias in determining the lowest $\chi^2$ values used in the final stage.
%%%%%%%%%%%%%%%%%%%%%%%%%%%

\subsection{Training}
\label{sec2B}
In the next step we select PDFs from the envelope to: {\it i)}  generate training data, or the code vectors; {\it ii)}  place vectors on the map (map vectors).
%%%%
%%%%
The training proceeds  through competitive learning which has each of the data vectors on the map competing with one other to be the best fit for a given set of training data.

An input vector 
\[ \xi = [\xi_i^{(1)}, ..., \xi_i^{(n)} ] , \] 
 is presented to the map. $\xi$ is isomorphic to $V_i \equiv f_i$. 
The elements of $V_i$ and $\xi$ are each compared/matched onto one another (see Fig.\ref{SOM:fig2}) using a similarity metric. In our case the metric is the Euclidean norm, for given $k$-dimensional vectors $x$ and $y$, where  $x \rightarrow f^{code}_i(x_1,Q^2)... f_i^{code}(x_m,Q^2)$, a code vector is compared  to  $y \rightarrow f^{map}_i(x_1,Q^2)... f_i^{map}(x_m,Q^2)$, the map vectors, by calculating the difference between the two functions for each Bjorken $x$ value at the same $Q^2$, through 
\begin{equation}
\label{metric}
 L_2(x,y) =  \sqrt{ \sum_{j=1,k} \left( x_j- y_j  \right)^2}.  
\end{equation}
During an {\it iteration},  the entire set of code vectors is presented to the map vectors. The node whose map vector is most similar to the input vector (an input PDF in our case) is defined as the Best Matching Unit or (BMU), or  ``winning" PDF. 
The weights, $V_i$, of the BMU and of the surrounding nodes form a neighborhood, ${\cal N}$, of  a defined radius.  
The weights of the set of map vectors in ${\cal N}$ are then modified so that the map vectors move closer, according to the given metric, to the input vector.  
The way the vectors adjust their values is described by the following algorithm,
\begin{equation}
V_i(n+1) = V_i(n) +  h_{ci}(n) [\xi(n) - V_i(n)],
\label{learn_1}
\end{equation}
where $n$ is the iteration number, and $h_{ci(n)}$ is the {\it neighborhood function} defining the radius on the map which decreases with both $n$, and
the distance between the BMU and node $i$. In our case we use square maps of size $L_{MAP}$, and 
\begin{equation}
h_{ci}(n) = 1.5 \left(\frac{n_{train} -n}{n_{train}} \right) L_{MAP}
\label{learn_2}
\end{equation}
where $n_{train} $ is the total number of iterations.
After all nodes have been adjusted and the neighborhood radius has been reduced according to the  
rule specified in Eq.(\ref{learn_2}),  the training procedure is repeated.  
The unsupervised  part of SOMs training takes place as the cells that are closest to the BMU activate each other in order to ``learn" from $\xi$.  

\subsection{Mapping}
\label{sec2C}
Finally, we proceed to the mapping stage. 
Once the learning process is complete, each new set of PDFs will be associated with the location of
its BMU.   
The various map parameters that need to be fixed at this stage include the size and shape of the map, the number of iterations, the number of PDFs used in each training cycle, and the initial learning rate. The map parameters values are presented and discussed in detail in Ref.\cite{Ask}.
At the end of a properly trained SOM, cells that are topologically close to 
each other will contain PDFs which are similar to each other.
In the final phase clusters can emerge.  Note that the specific location of the clusters on
the map is not relevant and will not necessarily be the same from one run to another;
only the clustering properties are important.

The visualization associated with clustering in the previous step allows us to isolate the individual properties of the input data functions we use and create more accurate models of them (see Fig.\ref{ANN:fig3}). 
Our SOMs appear as two dimensional arrays.  
Since each map vector now
represents a class of similar objects, the SOM
is an ideal tool to
visualize high-dimensional data, by projecting it onto a low-dimensional map
clustered according to some desired similarity feature. 

\subsection{Genetic Algorithm}
\label{sec2D}
After the map is trained 
we use a  Genetic Algorithm (GA) whereupon we construct new envelopes which contain sets of PDFs that are generated from each previous iteration, and that are selected based on their $\chi^{2}$ values so that, after a number of  iterations we minimize it.  
The new map PDFs, or the input PDFs, are analyzed relative to known experimental data for the observable  values. 
The $\chi^{2}$ is evaluated for each (evolved) map cell; PDFs with the lowest $\chi^{2}$ values are used as seeds for the next set of input PDFs.  This process is repeated for $N$ iterations; over the course of these iterations the $\chi^{2}$ values eventually asymptotically approach a given value, which is referred to as the saturation value. Reaching the saturation values defines the ``stopping" criterion for the fitting procedure in our case.

%%%%%%% SECTION III
%%%%%%%
\section{Study of unpolarized DIS data}
\label{sec3}
The SOMPDF analysis has been applied so far to extract unpolarized PDFs from inclusive DIS type processes.  The general method was developed in Ref.\cite{Ask}.  
The approach described in \cite{Ask}, however, does not explicitly interpret either the clustering properties, or the pattern recognition features of SOMs. 
In order to explore the performance of SOMPDF for these aspects, as a case study, we consider the extraction of the $d/u$ ratio from the proton and neutron structure functions at $x \rightarrow 1$. 
A precise knowledge of the PDFs at high $x$ is important: the behavior of the PDFs in this region is largely undetermined in spite of the fact that several theoretical predictions can be made for $d/u$ at $x=1$. A more precise knowledge of both the $d$ quarks and the gluons distributions and of their correlations will give also essential information to compute the QCD cross sections for collider experiments (see {\it e.g.} discussion in Ref.\cite{Acc1,Arr,Holt}).
The large $x$ behavior of the proton's structure functions from inclusive scattering is difficult to track down since many possibly mutually interfering effects, ranging from Target Mass Corrections (TMCs), Large $x$ resummation (LxR), and Higher Twists (HTs), affect the extraction of the PDFs from data. 

This problematic makes the extraction of the $d/u$ ratio ideal  to be studied with the SOMPDF analysis. 
The ultimate goal of  this analysis will be to extend our method, including a full usage of the SOMs pattern recognition features,  to extract more complicated observables such as GPDs and TMDs  from data. 

We proceed by first showing how our approach makes use of  SOMs plus GA in a minimization procedure, thus producing a set of PDFs with errors. We subsequently focus on the map training and its clustering properties, exploring how the SOMs classification of input patterns can help identify the factors governing the high $x$ behavior. 

Differently from extractions performed at intermediate and small values of Bjorken $x$ where abundant
high-precision proton and deuteron structure function data exist, beyond $x \approx 0.5$ it becomes much harder to constrain the PDFs, in particular the $d$ quark distribution. 
The PDFs extraction is hampered by various uncertainties which introduce extra $Q^2$ dependences beyond PQCD evolution including Target Mass Corrections, Large $x$ resummation, and  possible flavor dependent higher twists \cite{BiaFanLiu}. For the  $d$ quarks furthermore, one has uncertainties associated with the
nuclear corrections in the extraction of neutron structure
function \cite{Acc1,Arr}. Recent data from the BONUS collaboration at Jefferson Lab reduced these uncertainties for $x \leq 0.7$ by using spectator tagging measurements \cite{Tka}. However, because of the limited range in $Q^2$
attainable at Jefferson Lab, the large $x$ data have low invariant mass, $W$, {\it i.e.}  they lie in the resonance region. The additional hypothesis of parton-hadron duality is therefore needed.     

The standard procedure in this situation is to assess  the impact of the various $Q^2$ dependent and nuclear effects on the extraction of the $d$-quark and gluon PDFs,  by a critical study in which these effects are considered one by one until a satisfactory definition of the uncertainty of the extraction is achieved.  
SOMs provide an ideal tool to optimize this procedure by unraveling, using their pattern recognition feature, the various trends in the data that lead to the $Q^2$ dependent effects as $x \rightarrow 1$. 
%%%
\subsection{Unpolarized SOMPDF set}
\label{sec3A}
The SOMPDF algorithm provides first of all a viable minimization procedure that allowed us to extract PDFs from experimental data. Both the extraction procedure and the associated error analysis are described in detail in Ref.\cite{Ask}. 

\begin{figure}
\includegraphics[width=8cm]{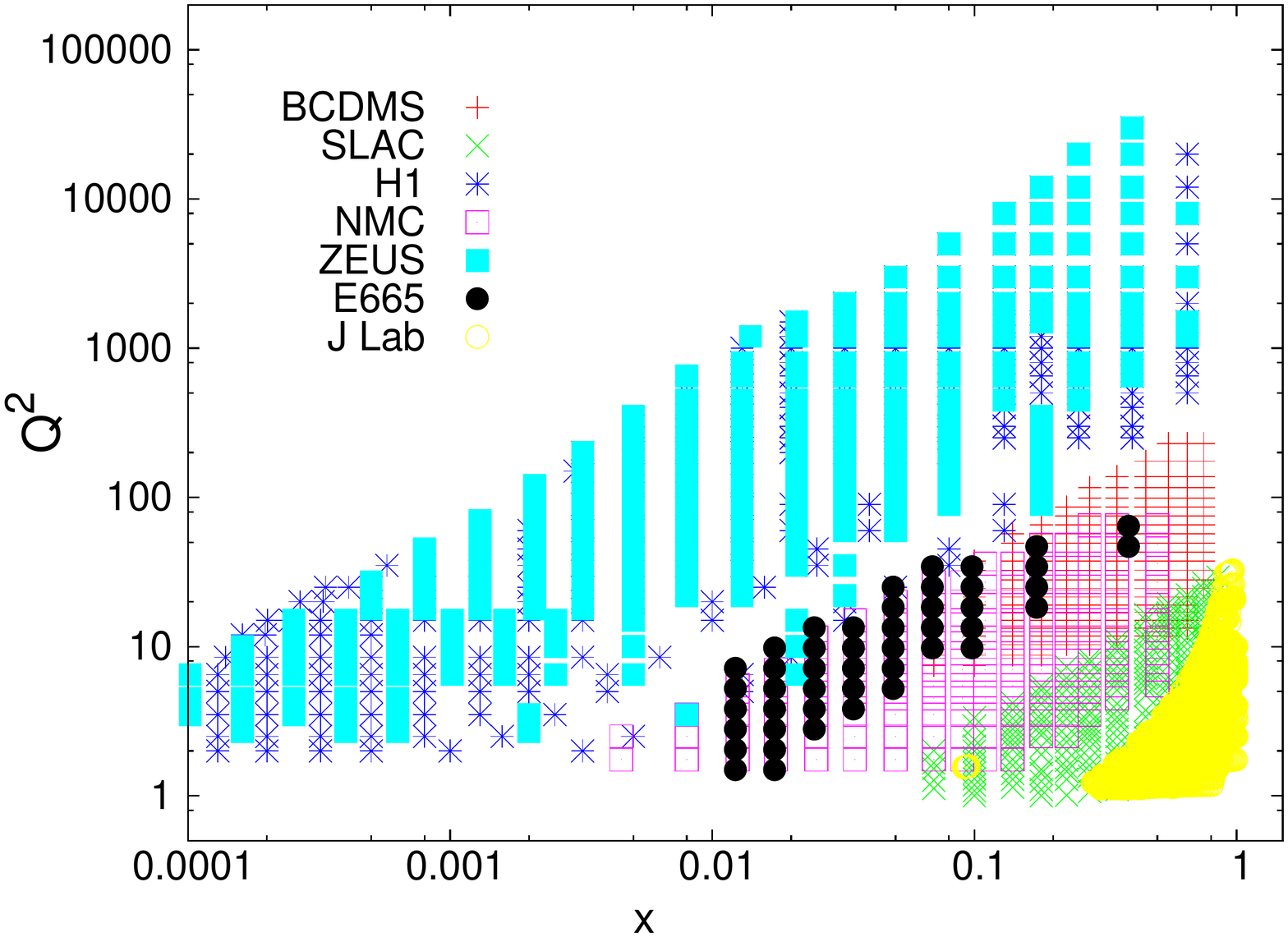}
\includegraphics[width=8cm]{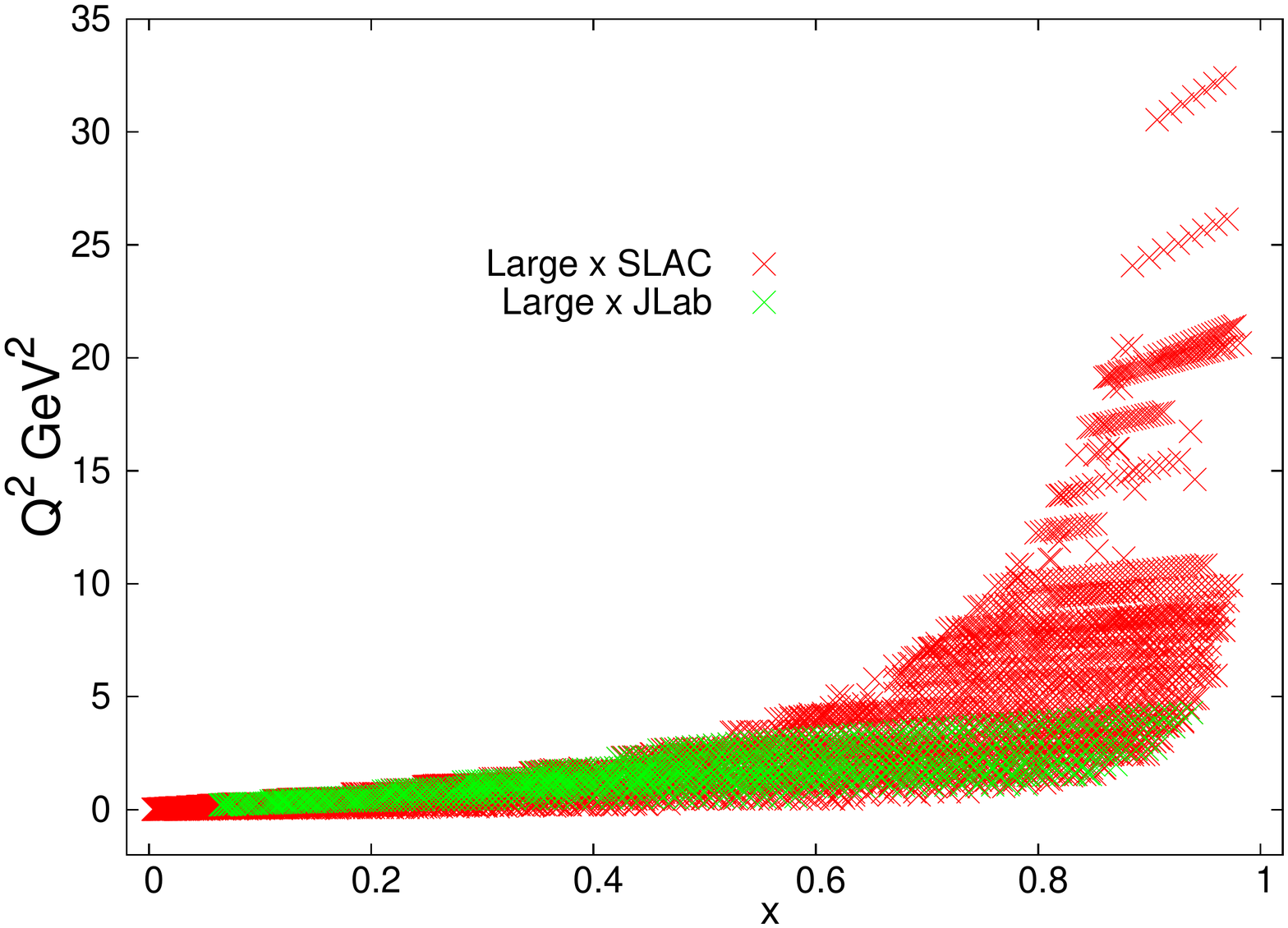}
\caption{(Color online) Kinematical range of the experimental data used in our analysis: (Left) \cite{SLAC,BCDMS,NMC,E665,H1,ZEUS,Jlab1}; (Right) large $x$ data used in this analysis 
\cite{SLAC,Jlab1}.}
\label{fig:kinematics}
\end{figure}

The set of data that we used in the present analysis is shown in Figure \ref{fig:kinematics}.
%%%% FIGURE data
%%%%%
%%%%%
The most recent set of PDF parameterizations as determined by various collaborations is listed in Ref.\cite{bench}.
Several of the analyses in Refs.\cite{bench} implement both DIS data and collider data. Similarly to the analyses in \cite{ABM,JR1,JR2}, we focus instead on  the most precise, or ``highest quality"  DIS data on lepton proton and lepton deuteron scattering from SLAC \cite{SLAC}, BCDMS \cite{BCDMS}, NMC \cite{NMC}, Fermilab E665 \cite{E665}, H1 \cite{H1}, ZEUS \cite{ZEUS} and Jefferson Lab \cite{Jlab1}. 
 At this stage of our analysis, with the aim of treating the large $x$ region,  this set of data allows test the working of our method in the cleanest way, avoiding the complications associated with the treatment of  different types of hard processes and inclusive observables. 

% In Table \ref{table:DIS} we list the data sets implemented in our analysis along with their kinematical range. 
%
%%%%

The starting point of our approach is the construction of the PDFs envelope which defines  the SOMPDFs initialization procedure (see Section \ref{sec2A}). 
Once the maps are initialized, the training procedure begins. At the end of a training section the various PDFs are organized on the map.  

In Ref.\cite{Ask} we studied the number of map parameters that can be adjusted during the initialization and mapping stages. We found that the best choice of map parameters
guaranteeing  both an optimal speed of convergence, and a precise evaluation of clustering properties is as follows: 
$6 \times 6$ or $7 \times 7$ for the size of the SOM maps; 
 $n_{PDF}= 1-3$, for the number of PDF types to be used for mixing; 
 $n_{cell} =2$, for the number of  PDFs per cell;
$n_{gen}=4$, for the number of PDFs to be generated for each cycle during training;
$n_{NEW} =10$, for the number of new PDFs to be generated each cycle, 
$n_{step} =5$, for  the number of steps to be used in training each SOM;
 $L_R^0=1$, for  the initial learning rate; 
$N_{MAX}=250$, for the maximum number of GA iterations. 
Finally, the slope parameter based on the number of previous $\chi^2$ values to look at when checking whether the $\chi^2$ curve had flattened out yet, $s_{flat}= 2 \times 10^{-3}$.  

For the initialization procedure, the parameters of the PDFs were varied by multiplying with a normally-distributed random number with mean of $1$ and standard deviation of $1/10$ the magnitude of the original parameter value. A larger standard deviation of $4/10$ and $6/10$ was applied for greater variation but this has not resulted in the $\chi^{2}$ values dropping below $2.5$ after 200 iterations.

\begin{figure}
\includegraphics[width=9.cm]{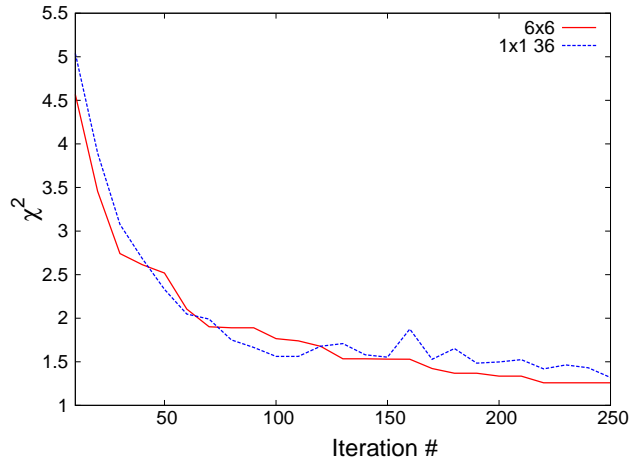}
%\hspace{-4.5cm}
\caption{(Color online) $\chi^2$ plotted vs. the  number of iterations for the GA. The two curves were obtained using a $6 \times 6$ maps and the GA only ($1 \times 1$ map).}
\label{chi1}
\end{figure}

We calculated the $\chi^2$ for the PDFs represented on the map (for each map cell)  following the definition in \cite{Pumplin1,Pumplin2},
\begin{equation}
\label{chi2}
\chi^2 = \sum_{i_{exp}} \chi^2_{i_{exp}} = \sum_{i_{exp}} \left[ \sum_{j_{data}} \left(\frac{{\cal N}_{i_{exp}} D_{j_{data}}^{i_{exp}} - T_{j_{data}}^{i_{exp}}}{\sigma_{j_{data}}^{i_{exp}} } \right)^2%
+ \left(\frac{1- {\cal N}_{i_{exp}}}{\sigma^{i_{exp}}}  \right) \right]
\end{equation}
where $\sigma^{i_{exp}}$ is the correlated error, in the normalization of the different data sets; ${\cal N}_{i_{exp}}$ (${\cal N}_{i_{exp}}=1$, for no offset of the normalization); $D_{j_{data}}^{i_{exp}}$, and  $T_{j_{data}}^{i_{exp}}$, refer to the data points (D) and theoretical estimate (T) at each given ($x$,$Q^2$); $\sigma_{j_{data}}^{i_{exp}}$ is the statistical uncertainty. 
Subsequent maps are run using the GA described in Section \ref{sec2B} namely, we populate the pool of randomized PDFs forming the envelope in each GA iteration with selected PDFs showing the best $\chi^2$ values from the previous iteration.  
As Fig.\ref{chi1} shows, at each GA iteration the $\chi^2$ decreases. We use the flattening of the $\chi^2$ as  a stopping criterium for our procedure.

%Although it is clear that SOMs provide a unique tool for detecting uncertain features of the data, we also remark that the SOM's clustering properties are an important component specifically in the minimization procedure.  The question of  whether the GA would be  sufficient on its own, or of whether it would be playing a dominant role, independently from the map architecture is in fact ruled out in our approach. This would be equivalent, in the NNPDF procedure,  to using only their Monte Carlo sampling, while eliminating the neural network as an interpolator to obtain the continuous PDFs. The whole ensemble of non linear correlations among the functions would be completely missed. In the minimization procedure this would generate solutions for local minima.    
Notice that the SOM's clustering properties are an important component even when considering only the minimization procedure, that is even before specifically searching for patterns in the data.   The GA would not be  sufficient on its own because it misses the SOM intermediate step (that can be compared to the supervised ANN intermediate layers where the weight fixing takes place) in which the PDFs that get represented on the map cluster according to latent variables/non-linear statistical connections. Even if a minimum $\chi^2$ is reached using a GA algorithm only, since important statistical connections are missed, one cannot rule out that the solutions obtained in this case might only represent a local minimum.
Nevertheless to check this point a quantitative analysis was performed in Ref.\cite{Ask}, that confirm that the map clustering property improves -- even if by small amounts --  the efficiency of the fit, or the trend and speed with which the $\chi^2$ is minimized.   The $\chi^2$ obtained in the two cases, {\it i.e.} using a $6 \times 6$ map, and the only the GA, equivalent to iterating 36 times a $1 \times 1$ map,  are shown in Figure \ref{chi1} plotted vs. the number of iterations of the GA. 
%%% CHI SQUARED FIGURE
%%%%%%%%%

For the error analysis we used the Lagrange Multipliers method. This is most indicated for SOMs because it uses directly the observables, therefore making it possible to circumvent the fact that we have no analytic forms with tunable parameters. The application of the Lagrange Multiplier method to PDFs global analyses was outlined in Ref.\cite{Pumplin1}. In Ref.\cite{Ask} we give a detailed description of our approach.
This method evaluates the variation of the $\chi^2$  along a specific direction defined by the maximum variation of a given physical variable. 
In our case the physical variables are the proton (deuteron) structure functions $F_2^{p(d)}$.
We take $F_2^{p(d)}$  as
determined by the SOMPDF code along with their  $\chi^2$ values
calculated from the comparison with the experimental data, $F_2^{exp}$. We then define an 
interval for the Lagrange multipliers, $\lambda \in [-200,200]$, and we increase $\lambda$
in increments of $10$. We generate sets of 
``pseudo experimental data" by shifting $F_2^{p(d)}$ for each given $x$ and $Q^2$ bin
by $\pm \Delta F$, and we repeat the SOMPDF fit for each new data set.
The new structure functions are defined by a corresponding set 
of new individual PDFs, $F_2^{exp, \, NEW}$. 
The difference between the individual 
PDFs from the limiting upper and lower $F_2^{exp, \, NEW}$ values define
then the Lagrange error for each of the individual PDFs for the original $F_2^{p(d)}$.

The uncertainties obtained with the Lagrange multipliers method, are not only more physically sensible, but they also differ quantitatively from a simple statistical analysis applied to the joint set of final envelope PDFs and map PDFs (the uncertainty is smaller and asymmetric)  .

%%%%%%%%%%
%%%% FIGURE MAPs
\begin{figure}
\includegraphics[width=9.cm]{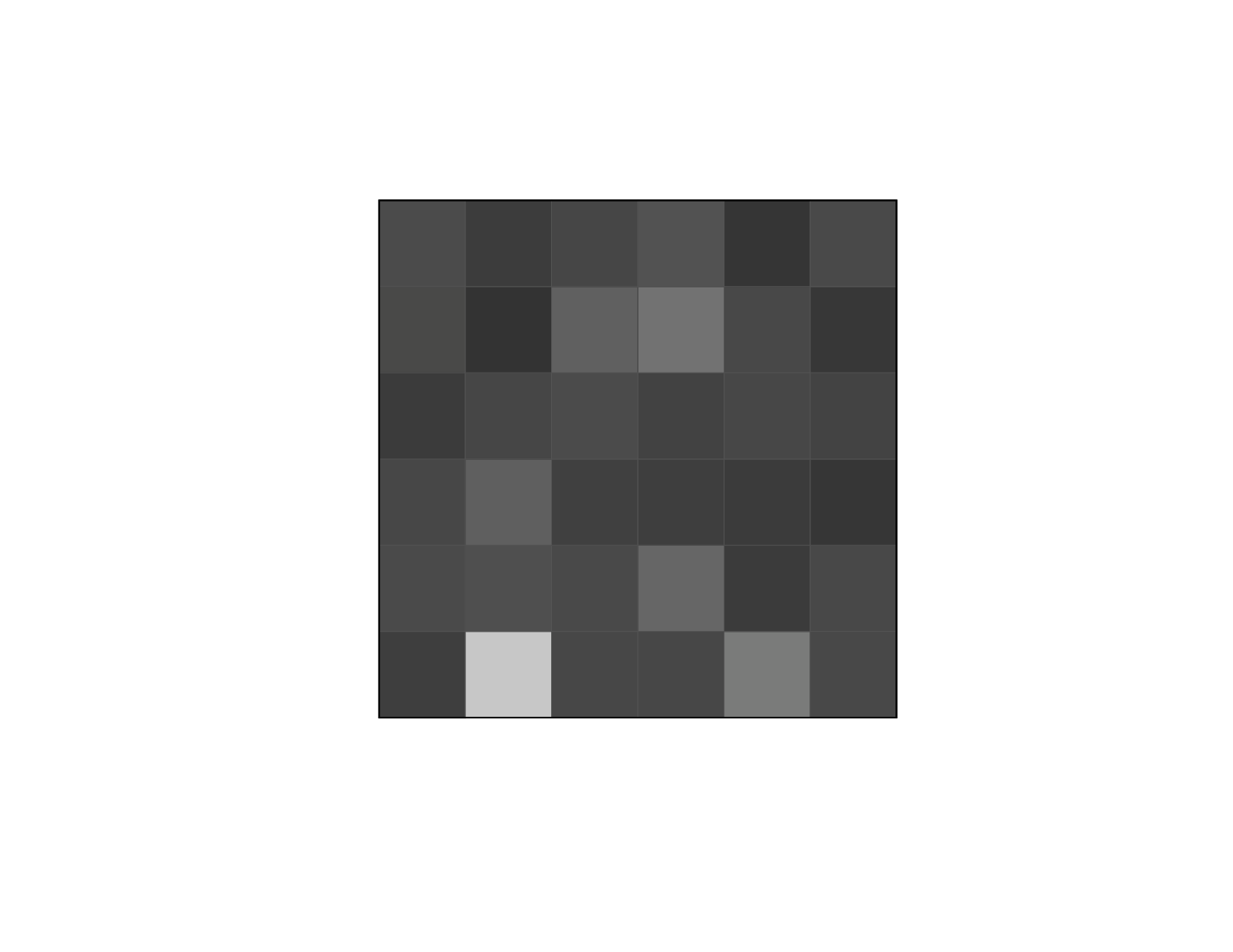}
\hspace{-1.2cm}
\includegraphics[width=9.cm]{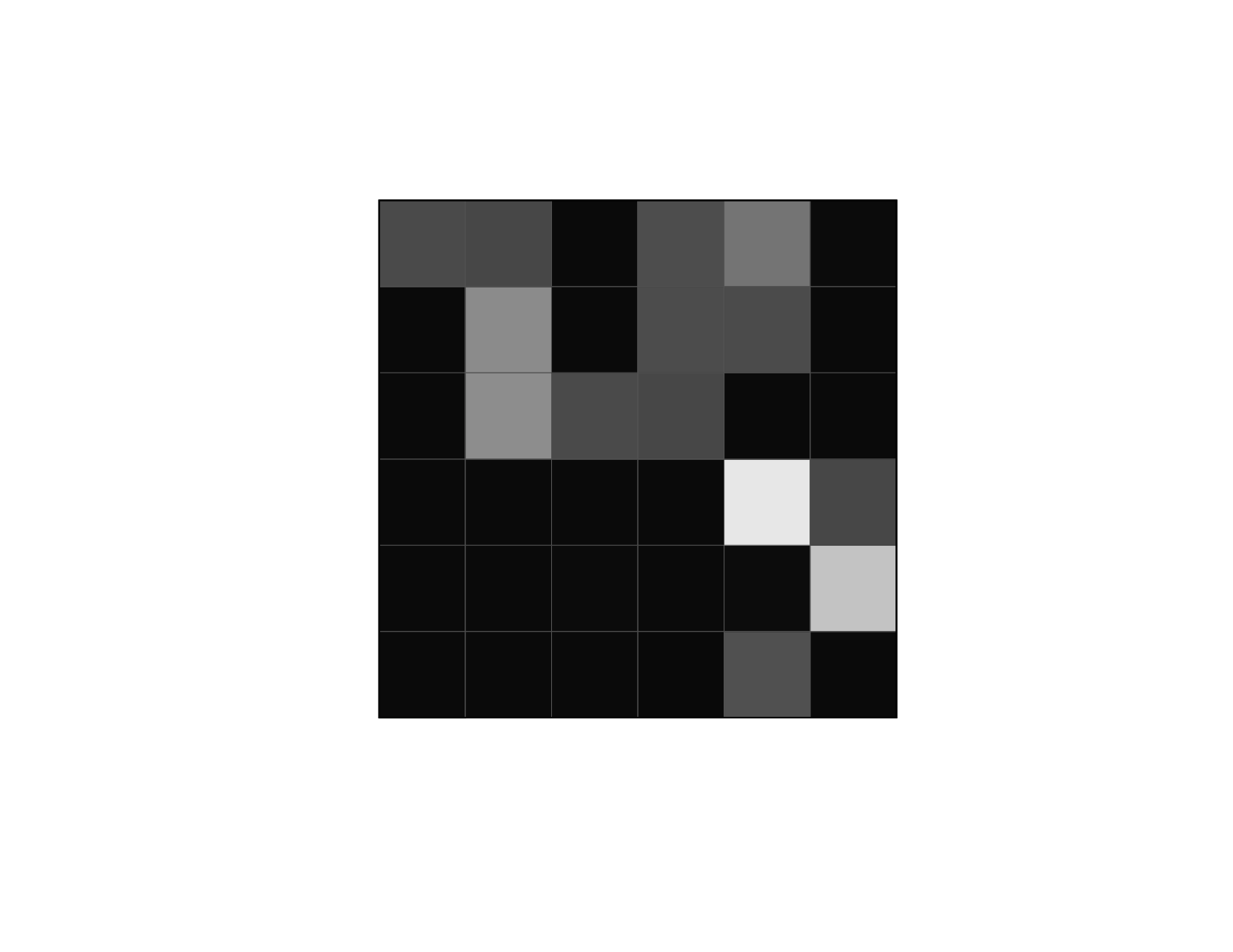}
\caption{Values of $\chi^2$ represented on $6 \times 6$ maps for the initial GA iteration (left) and  the final  one {\it i.e.} after $N=250$ iterations (right)  . The values of $\chi^2$ are lowest for the darker squares. One can see both the working of the clustering property of the map (PDFs with similar $\chi^2$ are located in neighboring map cells), and the GA (the $\chi^2$ is sensibly lower  .}
\label{chimap1}
\end{figure}
%%%%%%%
PQCD evolution at NLO was taken into account in order to compare all the envelope PDFs to data given at  specific $Q^2$ values.
%%%%
%%%% alpha_S
%%%%
Notice, in this respect, that the value of $\alpha_S(M_Z)$ which was allowed to vary in the range  $\alpha_S(M_Z)= 0.1135-0.1195$ consistently with other PDF extractions. 
The correlation between $\alpha_s$ and the PDFs uncertainty \cite{CT_alpha} is implicitly taken into account in our approach. 
%A detailed study will be presented in \cite{Aska_prep}.  

\begin{figure}
\includegraphics[width=8.5cm]{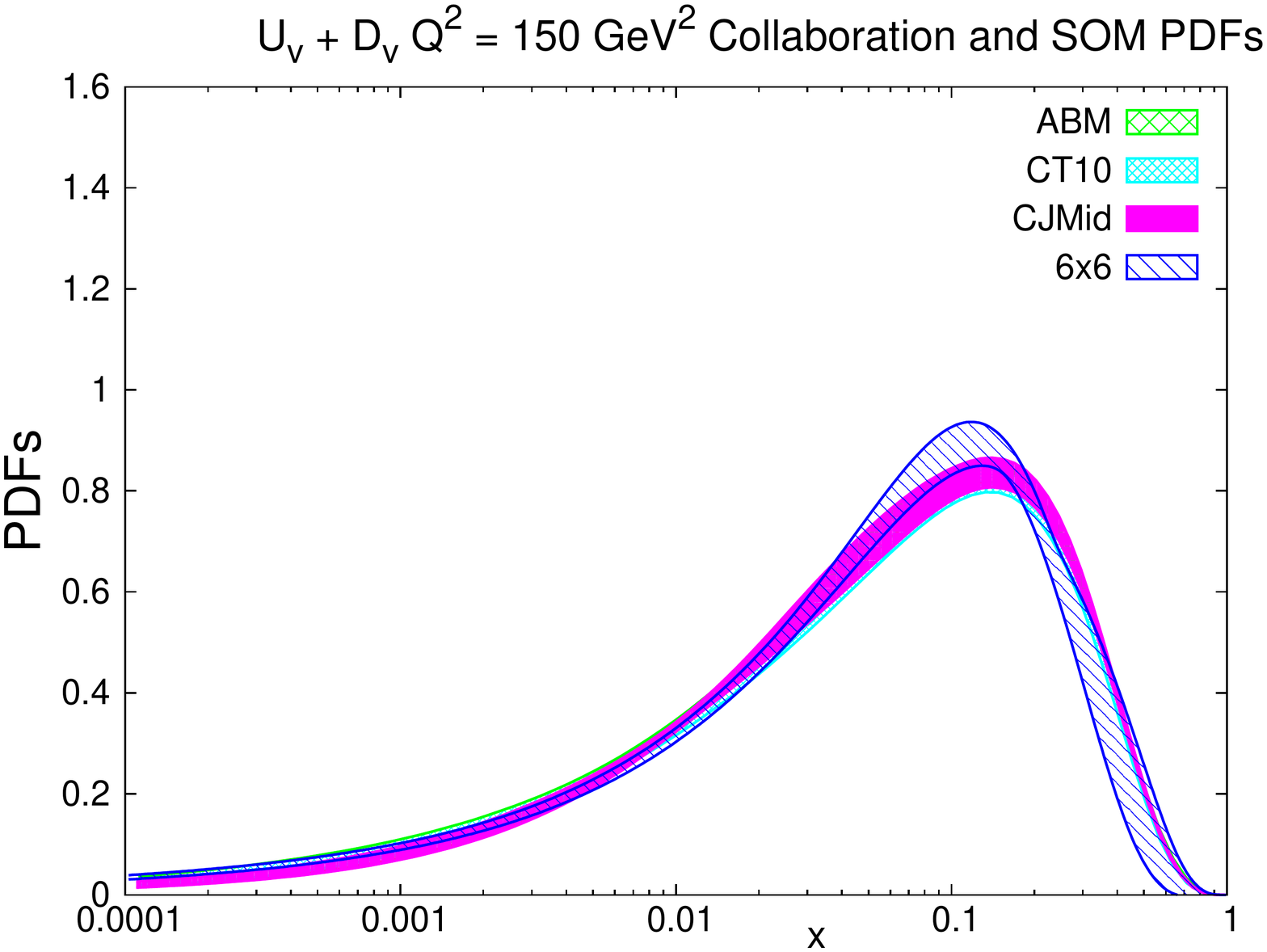}
\includegraphics[width=8.5cm]{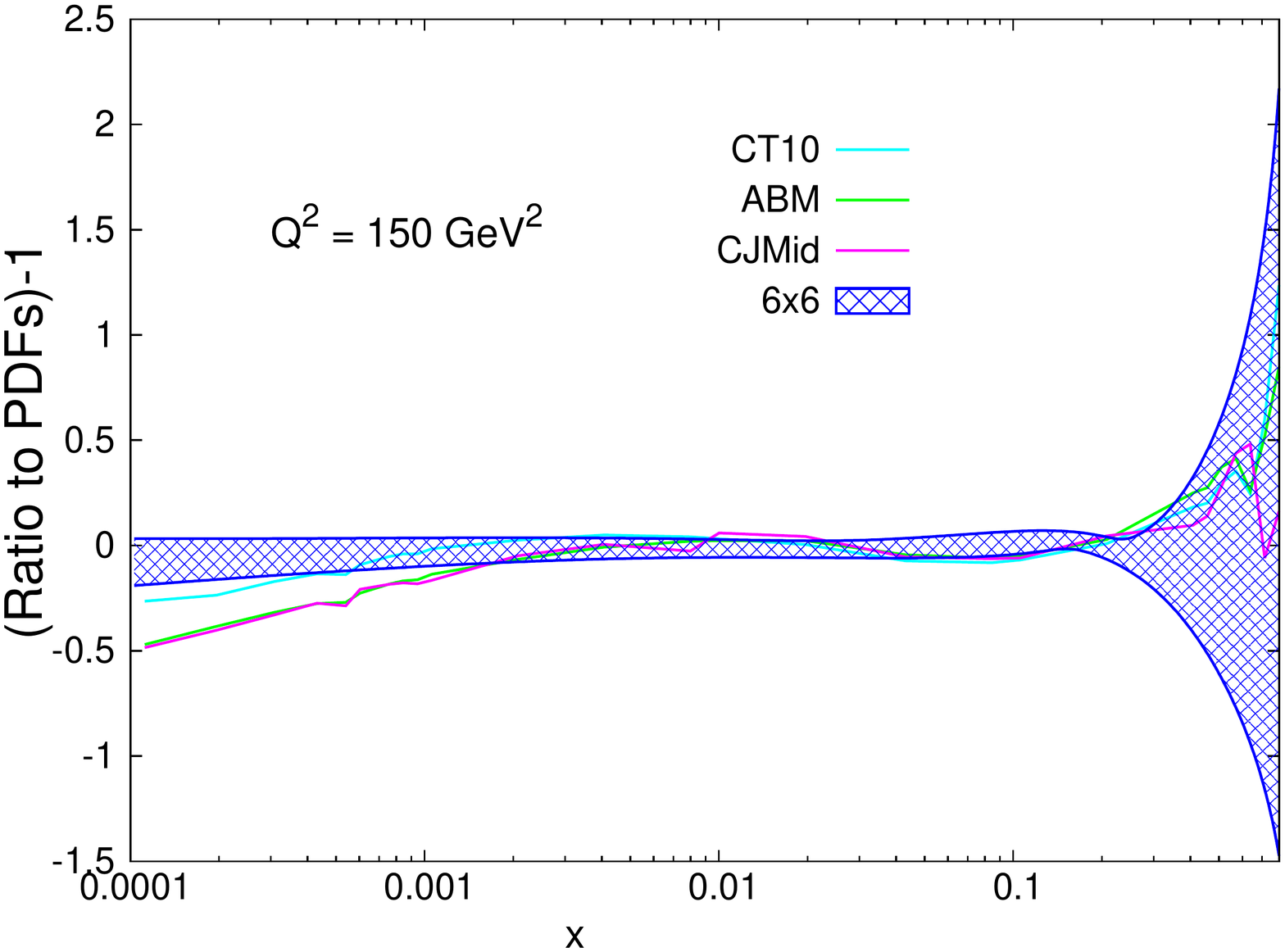}
\caption{Color online) Left: NLO $u_v+d_v$ at $Q^2=150$ GeV$^2$ calculated with a $6 \times 6$ map, using 250 GA iterations. The uncertainty band was evaluated using the Lagrange multipliers method described in the text. Our results are compared with NLO results from \cite{CT10,ABM}; Right: the pull, $(u_v + d_v)_{SOM}/(u_v + d_v)_{Collab} - 1$, of the SOMPDFs calculated using the same 
PDFs from Fig.\ref{fig:uvdv} to the different collaboration results, CT \cite{CT10},  ABM \cite{ABM}, CJ \cite{Acc1,Acc2} at $Q^2=150$ GeV$^2$, labeled as $Collab$. The error band around 0 is the SOMPDF $\sigma$ uncertainty.}
\label{fig:uvdv}
\end{figure}

\begin{figure}
\includegraphics[width=8.cm]{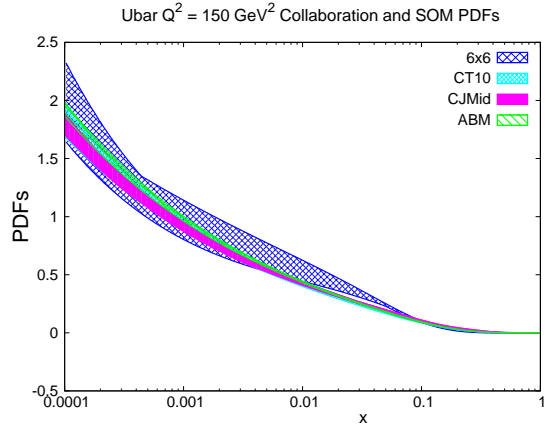}
\caption{(Color online) Same as Fig.\ref{fig:uvdv} (Left) for the $\bar{u}$  component.} 
\label{fig:ubar}
\end{figure}

\begin{figure}
\includegraphics[width=8.cm]{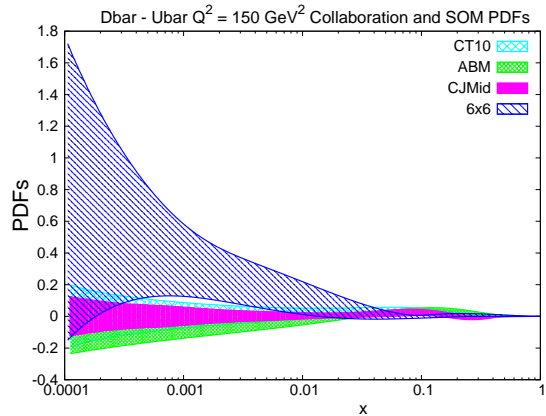}
\caption{(Color online) Same as Fig.\ref{fig:uvdv} (Left) for $\bar{d}-\bar{u}$.} 
\label{fig:ubdb}
\end{figure}

\begin{figure}
\includegraphics[width=8.cm]{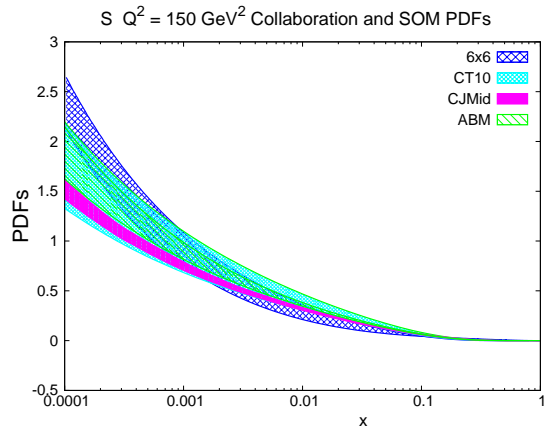}
\caption{(Color online) Same as Fig.\ref{fig:uvdv} (Left) for the $s$ quarks distribution.}
\label{fig:s}
\end{figure}

\begin{figure}
\includegraphics[width=8.cm]{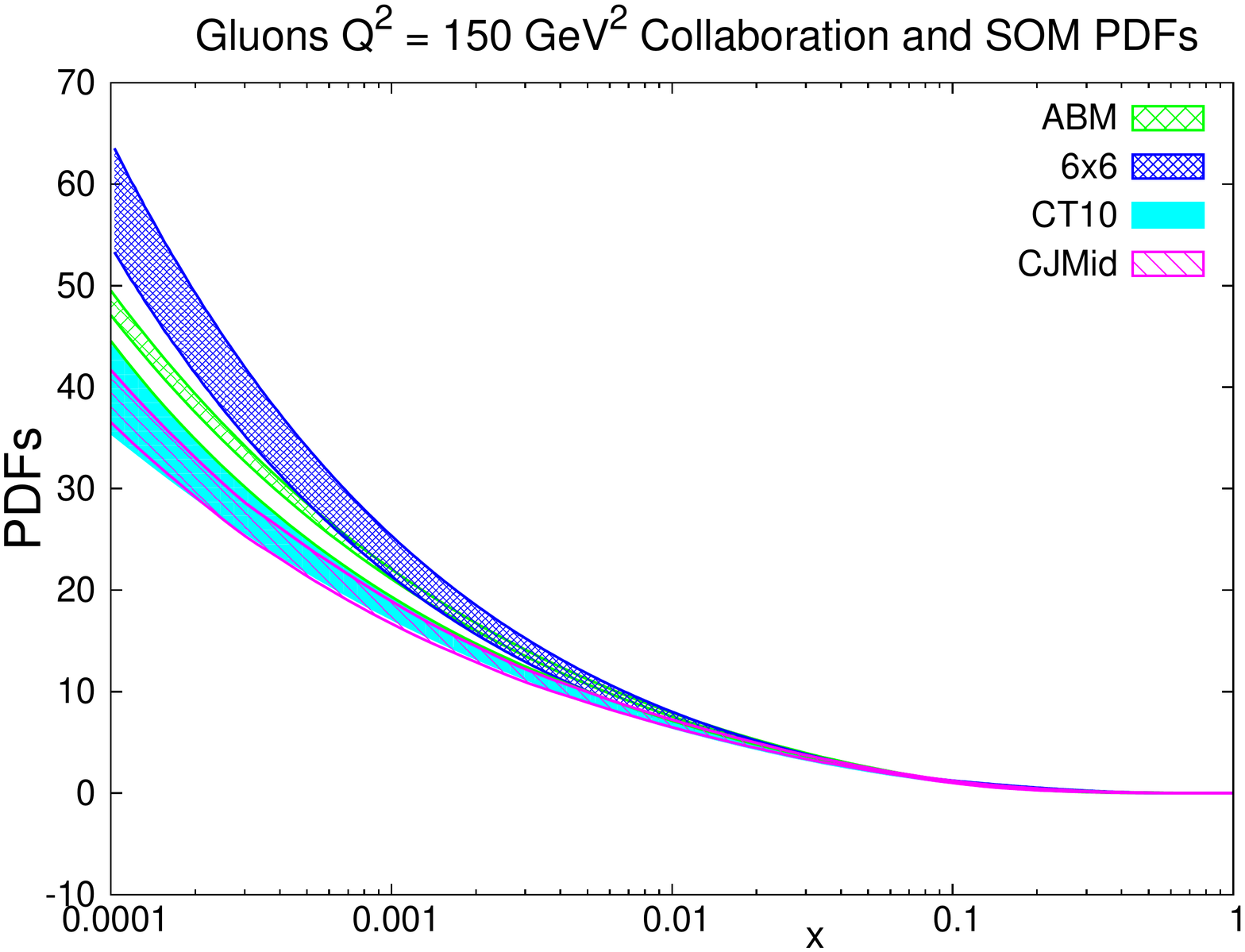}
\includegraphics[width=8.cm]{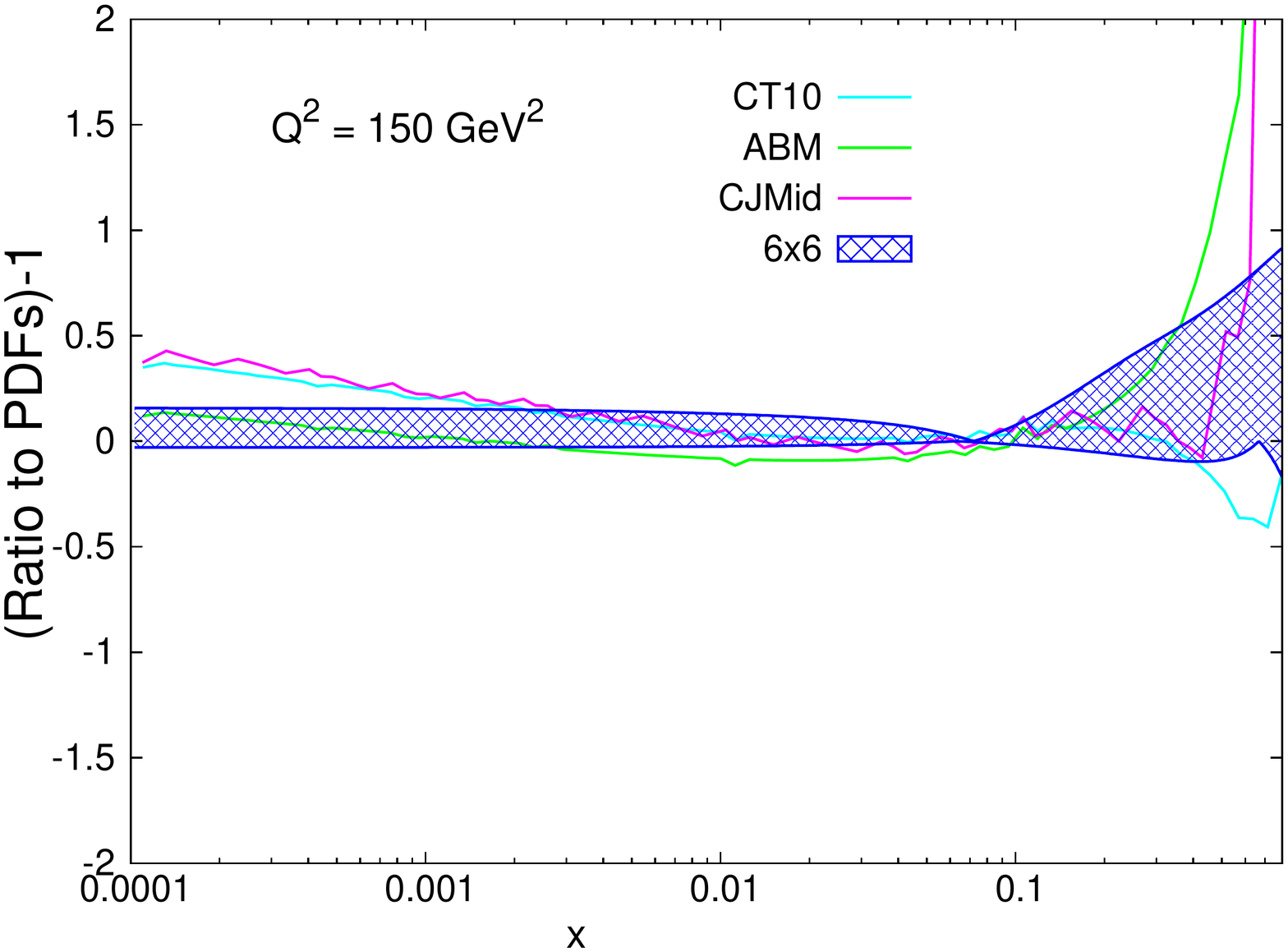}
\caption{(Color online) Same as Fig.\ref{fig:uvdv} for gluons.}
\label{fig:gluon}
\end{figure}

The PDFs extracted with the SOMPDF algorithm are presented in Figures  \ref{fig:uvdv}, \ref{fig:ubar}, \ref{fig:ubdb}, \ref{fig:s}, \ref{fig:gluon}. In each figure we compare our analysis to results from various collaborations: CT10 \cite{CT10}, ABM \cite{ABM}, and CJ  \cite{Ball.2010} at $Q^2=150$.  The error bands were calculated using  the Lagrange multipliers method (see \cite{Ask} for details).  Results obtained using the GA only, skipping the map construction, are also shown in the figures (labeled as $1\times1$). In Fig.\ref{fig:uvdv} and Fig.\ref{fig:gluon} we also show the pull ratio, $R_{pull}= (u_v + d_v)_{SOM}/(u_v + d_v)_{Collab} - 1$, of the SOMPDFs calculated using the same 
PDFs from Fig.\ref{fig:uvdv} to the different collaboration results, CT \cite{CT10},  ABM \cite{ABM}, CJ \cite{Acc1,Acc2} at $Q^2=150$ GeV$^2$, labeled as $Collab$.

These results show that SOMPDF is a viable method which is capable of yielding quantitative results, consistent with current analyses on the extraction of PDFs from high energy inclusive data. 
  
%%%%% PDFS with errors FIGURES
%%% UV DV

%%%% Ubar 

%%%% Dbar Ubar 

%%%% S

%%%% gluons

%%%%% CLUSTERING MAPS d/u
\subsection{Training and Clustering Properties SOMPDF Determination of Large $x$ features}
\label{sec4}
The analysis we conducted so far represents a first necessary step which validates SOMPDF as a method to quantitatively analyze deep inelastic structure functions, and to extract the various PDFs from experimental data.  The SOM algorithm is flexible, and it can be applied similarly to a variety of processes and observables, from the same structure functions at $x \rightarrow 1$ to  deeply virtual exclusive and semi-inclusive processes. 
For these processes the pattern recognition features and clustering properties of the SOM algorithm turn out to be an important aspect  to both facilitate extracting information, or gaining insight into the different features of the data.

A recent thorough study by the CJ collaboration \cite{Acc1} points out that for an accurate extraction of the $d$ quark PDF  at large $x$, among the different processes and sets of data which have been explored recently for their sensitivity to this distribution, standard DIS on deuteron is still the most reliable source. The extraction therefore has an additional uncertainty from the nuclear corrections. Since the $u$ quark at large $x$ is largely obtained from proton data, and it is therefore not affected by nuclear corrections, an efficient way of representing the possible variations in the $d$ quark distribution is through the ratio $d/u$. 
At leading order (with no flavor mixing in the higher twist terms), keeping  only the $u$ and $d$ distributions which are dominant at large $x$, one can get the ratio from the observables as,
\begin{equation}
\frac{d}{u} = \frac{4 \, \displaystyle\frac{F_2^n}{F_2^p} - 1}{4 - \displaystyle \frac{F_2^n}{F_2^p} } 
\end{equation} 
Besides facilitating the extraction of the large $x$ $d$ quark distribution,  the value of $d/u$ at $x \rightarrow 1$ is an interesting theoretical quantity in its own merit since it is one of the cleanest examples where non-perturbative quantum-field-theory-based/QCD methods -- Lattice QCD, Generalized Parton Distributions, and Dyson-Schwinger Equations approaches  -- and the various non-perturbative low energy QCD models for the flavor and spin structure of the nucleon can provide distinctive predictions (see {\it e.g.}  \cite{HoltRob} for a review). 

Our analysis includes data at larger values of $x$ by extending the kinematical range of the final state invariant mass, $W^2$, to $W^2\gtrsim 2 $ GeV$^2$  \cite{Acc3}, {\it i.e.} into the nucleon resonance region. To obtain smooth curves in the resonance region, we developed a method that interpolates the data with Bernstein polynomials \cite{AskLiu}.  This allows us to include in the analysis bins at very large $x$ ($x>0.9$) for a range of $Q^2$ values in the multi-GeV region.

\begin{figure}
\includegraphics[width=8.cm]{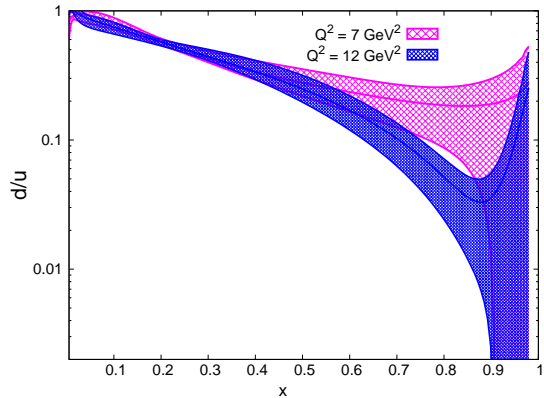}
\caption{(Color online) $d/u$ ratio extracted using data in two different regions of $Q^2$: $Q^2= 12$ GeV$^2$ \cite{ArrCoe}, and $Q^2=7$ GeV$^2$ \cite{Acc3}. The latter are dominated by nucleon resonances starting from $x\approx 0.7$. }
\label{fig:doveru}
\end{figure}

The goal of the SOMPDF analysis is to simultaneously take into account, compare, and study the correlations among various theoretical contributions that affect the extraction of PDFs at large $x$: Target Mass Corrections (TMCs), Large $x$ resummation effects (LxR), and nuclear corrections. 
As a first result in Figure \ref{fig:doveru} we present our extraction  in two different $Q^2$ regions, namely $Q^2 \gtrsim 12$ GeV$^2$, which is not affected by resonances up to $x \approx 0.85$ (the experimental data were taken from \cite{ArrCoe}),  and $Q^2=7 $ GeV$^2$, where the resonance structure becomes important at  $x \gtrsim 0.7$. 
The results presented in the figure were obtained by applying only the nuclear corrections with variable $x$ and $Q^2$ dependent smearing factors, $S_{p(n)}$ such that, 
\begin{equation}
F_2^d(x,Q^2)  = \frac{1}{2} \left(S_p(x,Q^2) F_2^p(x,Q^2) +  S_n(x,Q^2) F_2^n(x,Q^2) \right)
\end{equation}
%%%% d/u
%%%%%

From the figure one can see that both TMCs and LxR produce an effect on the $Q^2$ dependence, {\it i.e.} they do not seem to cancel in the ratio. 
An important point of the analysis is the error determination. The data at $Q^2=7$ GeV$^2$ lie in the resonance region at large $x$, leading to a less precise determination of the structure function with respect to the $Q^2=12$ GeV$^2$ data. In addition to that, both regions are largely affected by nuclear smearing.  These, namely the larger errors in the resonance region and the presence of the in principle unknown smearing factors, are the reasons why in the $x \rightarrow 1$ limit, {\it i.e.} for $x \geq 0.9$, despite there are no bins at $Q^2=12$ GeV$^2$, while we  introduced more bins  at $Q^2=7$ GeV$^2$, the errors become both comparable to each other and large. 

The purpose of Fig.\ref{fig:doveru} is indeed to illustrate how the various components of the analysis contribute to make the uncertainty large as $x\rightarrow 1$.   
The subsequent  steps of the analysis involve introducing all corrections: TMCs, LxR and nuclear effects. These can be considered either individually, introducing them one by one, as well as simultaneously. However, even after completion of a full analysis, and after extending the set of data to the largest $x$ available, the uncertainty remains large due to the fact that it is very taxing to pin down all the various effects, their correlations, and the correlations among the PDFs.     
The most visible correlations were already pointed out in the pioneering work in Ref.\cite{Ale}. A more accurate analysis was more recently performed in Ref.\cite{Acc1} where the correlation between the $d$ quark distribution and nuclear effects was studied in detail. There it was also observed that despite the interference between nuclear effects and the size of the $d$ quark PDF is rather predictable, this  carries over indirectly, essentially through $Q^2$ evolution,  to the large $x$ gluon PDF. 

This situation calls for a different approach. An approach that will allow us to discern patterns in complex data sets.   In upcoming work using the SOMPDF method we explore how to discriminate more efficiently among different models, and among different features of models. More specific work in the large $x$ sector, where many effects such as target mass corrections, large $x$ resummation, higher twists, and nuclear dynamics corrections are simultaneously present, is in preparation \cite{AskLiu}.

%%%%%
%%%%% SECTION IV CONCLUSIONS
\section{Conclusions and Outlook}
\label{sec5}
We presented an overview of a new method to extract PDFs from hard scattering processes.  PDFs analyses are complex notwithstanding the large amount of experimental data that have been collected throughout the years, and the fact that theoretical developments in this sector can be considered in a mature stage.   
The main challenge is to separate in a statistically significant way the various theoretical components by which PDFs are embedded into proton's structure. 

Recently, SOMs, a neural network that is based on an unsupervised learning algorithm has been proposed as a tool to explore classification patterns in high energy physics problems, including jet identification, and PDFs fitting (SOMPDF). 
In this respect, similarly to NNPDF,  SOMPDF  provides parameterizations that are not biased by the choice of a specific functional form for the PDFs.
The way a generic neural network  approach functions is by  increasing the flexibility of the functional form used to fit the data by handling a much larger number of parameters  given by the  network's weights. The price to pay for the increased flexibility and unbiasedness is given by the intractability of kinematical regions where experimental data that are necessary to train the network are scarce. Through the self-organizing procedure, or  the ability to create its own classification of the input data, SOMs achieve a greater predictive power compared to generic ANN-based fits. At the same time,  since SOM works by exploiting latent correlations existing among data SOMPDF can be used to explore patterns among PDFs and to pin down correlations both among different PDFs, and among PDFs and theoretical components 

In this paper we showed how the SOMPDF method works quantitatively. We illustrated the features of a parameterization for PDFs at NLO.  The error  analysis which was carried out using the Lagrange multipliers method. Our method can parallel and support the ANN based effort. 

%%% Outlook
As a case study, we illustrated the specific issues that arise in the study of PDFs such at large Bjorken $x$. 
Future applications of our method will include its application 
to hard  exclusive scattering experiments.  
These will include, for instance, both recent and forthcoming measurements at Jefferson Lab and at CERN (COMPASS) which are aimed at disentangling eight Generalized Parton Distributions (GPDs) depending on two extra kinematical variables (these allow us to describe transverse spatial partonic configurations). 
An even larger number of higher twist distributions will also play an important role in the interpretation of experimental data. 
GPDs analyses clearly involve many more degrees of freedom, and therefore a higher degree of complexity in the treatment of the hadronic structure.
This makes it a daunting task to perform fully quantitative fits including a precise treatment of the uncertainty.
SOMPDFs provide an ideal tool for analyzing these complex observables owing to the various features that were discussed in this paper. Most importantly, differently from a standard neural network they will allow us to extrapolate to regions where data are  scarce due to the unsupervised learning algorithm. 

We conclude that SOMs due to their unique capability of handling correlations which appear in the analysis of nucleon structure provide a promising tool to obtain enhanced information on the new complex, multivariable dependent phenomena explored in hadronic physics.

\acknowledgements
We thank Donal Day for support and discussions about data sets, and John Arrington for providing us with details of his analysis of PDFs at large Bjorken $x$. We are also indebted to UVACSE for assistance with code development and resource utilization. 
This work was supported by the U.S. Department
of Energy grants DE-FG02-01ER4120 (E.M.A. and  S.L.), and DE-FG02-96ER40950 (E.M.A.).
%%%%%%%%%%%%%%%%%%%%%%%%%%%%%%%%%%%%%%%%%%%%%%%%%%%
%%%%%%%%%%%%%%%%%%%%%%%%%%%%%%%%%%%%%%%%%%%%%%%%%%%
%%%%%%%%%%%%%%%%%%%%%%%%%%%%%%%%%%%%%%%%%%%%%%%%%%%

\end{document}